\documentclass[11pt,twoside]{article} 
\usepackage{cspm-asp2006}
\usepackage{epsfig,graphicx,natbib,url}  
\usepackage{lscape} 
\begin{document}

\setcounter{page}{605}

\markboth{Balthasar et al.}{GREGOR} 
\title{GREGOR: the New German Solar Telescope} 
\author{H.~Balthasar$^1$, 
        O.~von~der~L\"uhe$^2$, 
        F.~Kneer$^3$, 
        J.~Staude$^1$,
        R.~Volkmer$^2$, 
        T.~Berkefeld$^2$, 
        P.~Caligari$^2$, 
        M.~Collados$^5$, 
        C.~Halbgewachs$^2$, 
        F.~Heidecke$^2$, 
        A.~Hofmann$^1$, 
        M.~Klva\v na$^4$, 
        H.~Nicklas$^3$, 
        E.~Popow$^1$, 
        K.~Puschmann$^3$, 
        W.~Schmidt$^2$, 
        M.~Sobotka$^4$, 
        D.~Soltau$^2$, 
        K.~Strassmeier$^1$ and A.~Wittmann$^3$  } 
\affil{$^1$ Astrophysikalisches Institut Potsdam, Germany, \\
       $^2$ Kiepenheuer-Institut f\"ur Sonnenphysik, Freiburg, Germany, \\
       $^3$ Institut f\"ur Astrophysik, Universit\"at G\"ottingen, Germany, \\
       $^4$ Astronomical Institute AS, Ond\v rejov, Czech Republic, \\
       $^5$ Instituto de Astrof\'\i sica de Canarias, La~Laguna, Spain} 

\begin{abstract}  
 GREGOR is a new open solar telescope with an aperture of 1.5~m.  It
replaces the former 45-cm Gregory Coud\'e telescope on the Canary
island Tenerife. The optical concept is that of a double Gregory
system. The main and the elliptical mirrors are made from a
silicon-carbide material with high thermal conductivity. This is
important to keep the mirrors on the ambient temperature avoiding
local turbulence. GREGOR will be equipped with an adaptive optics
system. The new telescope will be ready for operation in
2008. Post-focus instruments in the first stage will be a spectrograph
for polarimetry in the near infrared and a 2-dimensional spectrometer
based on Fabry-P\'erot interferometers for the visible.
 
\end{abstract}

\section{Introduction}

Magnetic features on the sun are rather small or have an internal fine
structure with typical scales of the order of 50 -- 100~km.
Variations occur on short time scales of a few minutes.  To collect
the required number of photons, large telescopes are needed to
investigate such features.  Such a telescope must have good
polarimetric properties in order to enable high precision polarimetric
measurements.  A German consortium consisting of the
Kiepenheuer-Institut f\"ur Sonnenphysik, the Institut f\"ur
Astrophysik in G\"ottingen and the Astrophysikali\-sches Institut
Potsdam presently builds a new solar telescope with an aperture of
1.5~m, GREGOR.  International partners in the construction are the
Astronomical Institute Ond\v rejov (Czech Republic) and the Instituto
de Astrof\'\i sica de Canarias (Tenerife, Spain).  GREGOR replaces the
previous 45-cm Gregory-Coud\'e-telescope on Tenerife.  Reports on the
progress in constructing this telescope can be found in
\citet{horst-2005SPIE.5901...75V}, \citet{horst-2006SPIE.6267E..29V}
and \citet{horst-volkmer_goettingen}.

GREGOR is designed as an open telescope, so the air can flow freely through 
the telescope. Problems with plane-parallel entrance windows of the required 
size are avoided this way, but now one has to care for the absorbed heat
at the surfaces which are exposed to the full sunlight. Roughly 2000~W
of radiation fall on a primary mirror of 1.5~m diameter, and with a coating
of aluminum about 10\% will be absorbed. Therefore, active cooling of
the primary mirror is required to keep it at the ambient temperature.
If the material of the mirror body has a high thermal conductivity, it
is sufficient to blow cold air onto the backside of the mirror. This is
the reason why we decided not to use a standard glass-ceramic. Instead, the 
main mirror and the two following ones will be made from a silicon-carbide 
material. Besides its high thermal conductivity, this material has also 
the advantage of a quite high stiffness, so that the 
mirror is less bent due to the changing orientation during the day. 
In addition, the mirror mount is designed to minimize the influence of
gravitational forces on the mirror.
On the other hand it is very difficult to produce a mirror of this size 
fulfilling the optical requirements. More details can be found in 
\citet{horst-2006SPIE.6273E..22K}.

\begin{figure}
  \centering
  \includegraphics[width=\textwidth]{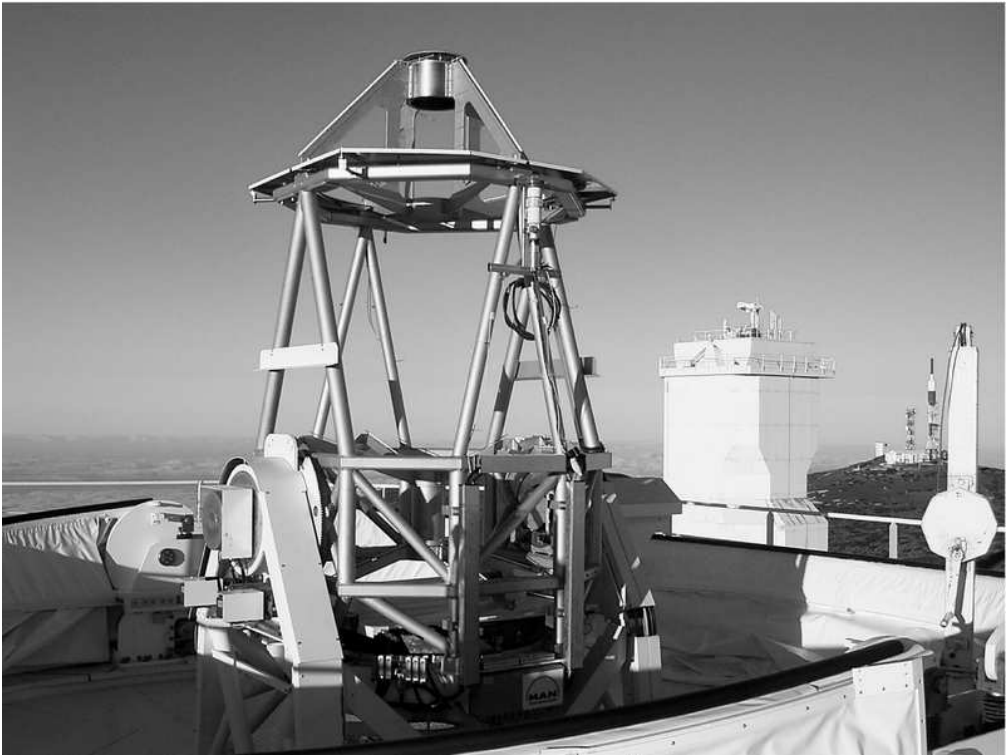}
  \caption[]{\label{hb-fig:structure}
  The mechanical structure of the telescope on top of the building. 
The dome is folded down. The Vacuum Tower Telescope is seen in the background.
}
\end{figure}

The telescope has an alt-azimuthal mounting, thus the telescope
structure is very compact and mechanical problems are reduced to a
minimum.  To realize the concept of an open telescope, and in order to
make room for the 1.5~m telescope, the dome of the GCT had to be
removed.  It was replaced by a foldable tent of the same type as that
of the Dutch Open Telescope on La~Palma. Since the erection of this
new dome several strong storms passed Tenerife which the dome
resisted.  The telescope structure was mounted at the site in 2004 and
is shown in Fig.~\ref{hb-fig:structure}.  The movements of the
telescope are computer-controlled guaranteeing a precise basic
guiding.  The fine tracking is done by an adaptive optics system.

The telescope is designed for the wavelength range 380\,nm through 12~$\mu$m.

The operation of the telescope and the control of the post-focus 
instruments will be done from a separate room in the third floor of the 
building. Graphical interfaces give access to all instruments and sensors.
 
\section{The Optical Path}

\begin{figure}
  \centering
  \includegraphics[width=13cm]{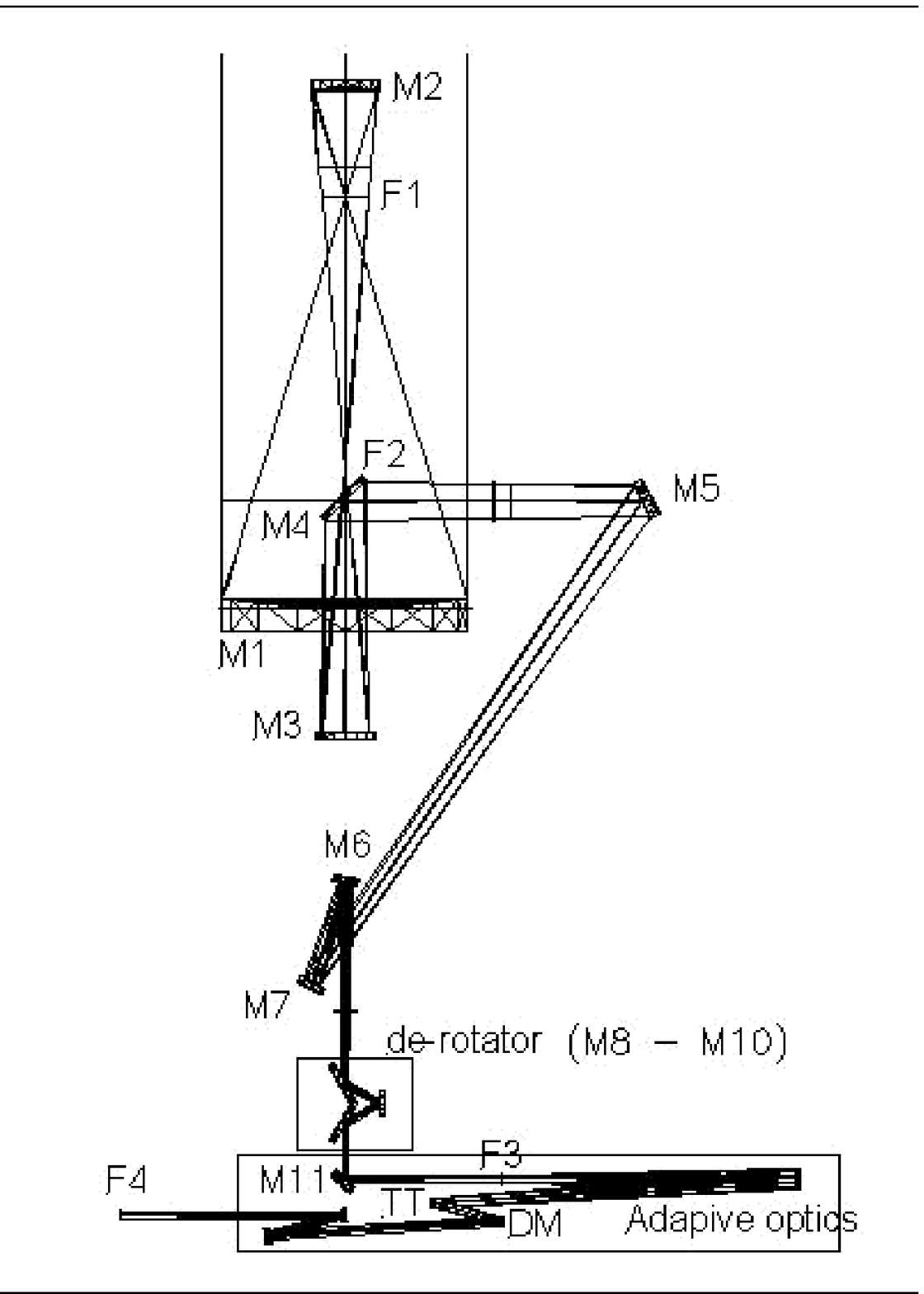}
  \caption[]{\label{hb-fig:optical_path}
  Optical scheme of the telescope.

}
\end{figure}

A scheme of the optical path is displayed in Fig.~\ref{hb-fig:optical_path}.
The optical concept is that of a double-Gregory Coud\'e telescope. 
This telescope has an effective focal length of 55~m.
The first three mirrors are aligned along the same optical axis to keep 
instrumental polarization to a minimum until the secondary focus. 

The primary mirror M1 has a diameter of 1.5~m, is shaped parabolically,
and has a focal width of 2.5~m. Cold air will be blown into pockets 
on the back side of the body of the mirror to keep temperature 
differences to the ambient below 0.3~K, sufficient to avoid heat-driven
air turbulence at the front surface of the mirror. 
The temperature of the cooling air is adjusted according to the prevailing 
air temperature.

At the primary focus F1 there is a field stop which allows a circular
field of 150 arcsec diameter to pass, alternatively a wider hole with
300 arcsec width can be inserted.  The rest of the light (more than
97\%!) is deflected out of the telescope.  Because of the concentrated
light at this location, this field stop needs active cooling by a
water circulation, which is realized in two circuits.  This way, the
temperature difference to the surrounding air will be kept below 5~K.

The secondary mirror M2 is an elliptical one with a diameter of 42~cm.
The two focal lengths are 67~cm and 2.3~m. This mirror will absorb
6~W, therefore passive cooling is enough to avoid temperature
differences of more than 2~K. The alignment of this mirror with
respect to M1 is extremely crucial for the optical performance of the
telescope. M2 will be mounted on a hexapod permitting the needed
degrees of freedom for movements. These movements will be done under
control of the adaptive optics system (AO).

Little space is available near the secondary focus F2, nevertheless 
there is the possibility to introduce a linear polarizer (a Marple-Hess
prism) and one of two achromatic quarter-waveplates (380 -- 800\,nm and
750 -- 1600\,nm, respectively). All these elements can be rotated with
a step width of 0.1 degrees allowing the instrumental calibration of
the instrument and a proper measurement of the circular polarization.
For alignments, also a grid target or a pinhole can be inserted here.
A more detailed report on polarimetry with GREGOR is given by 
\citet{horst-2006SUN.GEOS.AH}.

The tertiary mirror M3 is used to focus the telescope. It is again an
elliptical mirror with a diameter of 36~cm. The focal lengths of this 
mirror are 1.6~m and 10.1~m.
Passive cooling is sufficient for this mirror too.
 
M3 reflects the light beam to flat mirror M4 which is located at the 
intersection of the optical axis of the telescope and the elevation axis.
M4 reflects the light into the elevation axis. From there the beam  
passes three more flat mirrors (M5 -- M7), and is finally reflected into the 
azimuth axis of the telescope. 
Heat absorption is not a problem for these mirrors,
so they are made from ZERODUR. For the case that turbulence occurs in
this part of the optical path, glass plates are introduced between M4 
and M5 and after M7, and the tube inbetween can be evacuated. For 
observations at wavelengths longer than 2.5~$\mu$m, these plates must be 
removed.

A disadvantage of the alt-azimuthal mounting is the rotation of the final 
image. This will be compensated by a de-rotator consisting of three flat
mirrors M8 -- M10. In addition, the de-rotator can be used to change the 
orientation of the image with respect to the post-focus instruments, 
e.~g. the spectrograph slit. The de-rotator can be removed from the 
optical beam.

Mirror M11 reflects the light into the AO system. A filter wheel is 
placed at the tertiary focus F3; different field stops, a pinhole, 
a target, and a cover for dark exposures can be inserted here. Then the 
light passes the collimator M12, the tip-tilt mirror TT, the deformable 
mirror DM and the camera mirror M15, before mirror M16 sends the light 
to the requested post-focus instrument. M11 and M16 can be removed to 
bypass the AO. The deformable mirror DM has 80 electrodes and a free 
aperture of 55~mm. Each post-focus instrument will have its own wavefront
sensor of Shack-Hartmann type. The wavefront sensors will have 78
usable sub-apertures. One sub-aperture will be covered by 24 $\times$ 24
pixel, and one pixel will correspond to 0.5~arcsec. Further details about
the AO-system can be found in \citet{horst-2006SPIE.6267E..34S}.

\section{Post-focus Instruments}

In the first stage GREGOR will be equipped with a 2D-spectrometer and
a slit spectrograph together with an infrared polarimeter. Most post-focus 
instruments are placed in the optical laboratory in the fifth floor of
the building, just below the telescope. The entrance slit and the polarimetry 
unit of the spectrograph are placed here too, then the light is reflected  
down to the fourth floor where stable conditions are guaranteed for the 
spectrograph. 

The 2D-spectrometer is an upgrade of the G\"ottingen Fabry-P\'erot spectrometer
which was operated successfully for many years at the Vacuum Tower Telescope
(VTT). It is based on two new tunable Fabry-P\'erot interferometers (FPI) which
have a diameter of 70~mm. The FPIs are placed in the collimated beam near
the telescope pupil. The usable wavelength range is 530 -- 870\,nm, and the
resolving power is of the order of 250000.

The slit spectrograph of Czerny-Turner type is optimized for the near infrared 
to be used with
the Tenerife Infrared Polarimeter (TIP 2, \cite{horst-tip2}), which was operated at the VTT 
before. The heart of the spectrograph is the grating previously used at the GCT.
The focal lengths of both, collimator and camera mirrors of the new spectrograph 
are 6~m. 
At 1.1~$\mu$m it has a resolving power of 525000, and the dispersion corresponds to
52~pm/mm. 
The spectrograph can be used for any single wavelength range in the visible 
too, only the dispersion is not well adapted to the pixel size of modern 
CCD-detectors. A few combinations of two visible ranges to be observed 
simultaneously are possible. A detailed description of the spectrograph 
is given by \citet{horst-spectrograph}.

Our plans also foresee to move the Polarimetric Littrow Spectrograph (POLIS)
from the VTT to GREGOR as soon as GREGOR is ready. A description of this 
instrument was published by \citet{horst-2005A&A...437.1159B}. 
  
One free optical table is available for experimental setups or guest 
experiments. 

On a longer time scale, GREGOR will also be used for stellar astronomy, 
the Astrophysical Institute Potsdam plans a special spectrograph to
obtain stellar spectra over the whole visible range. This spectrograph 
will be placed in the fourth floor.

\section{Conclusion}

With its aperture of 1.5~m and equipped with an adaptive optics system, 
GREGOR will be the most powerful solar telescope for high resolution
spectroscopy in the range from 380\,nm through 2200\,nm for the near 
future. GREGOR has good polarimetric properties and will be equipped 
with a polarimeter for the infrared and the visible spectral range
(TIP and POLIS). 

First light and commissioning are expected to occur in 2008.\\

\acknowledgements 
This report benefits from the ESMN (European Solar Magnetism Network) 
supported by the European Commission under contract HPRN-CT-2002-00313.

\end{document}